\documentstyle[pra,aps]{revtex}

\begin{document} \title{Propagation of sound in two species Bose-Einstein condensates}
\author{D. Lima\thanks{email: dlima@ulb.ac.be}$^1$, P. Borckmans$^2$, G. Dewel$^2$\\
 $^1$Instituto de F\'{\i}sica - Universidade Federal da Bahia \\
 40210-340 Campus da Federa\c{c}\~ao - Salvador - Brazil \\
  $^2$ Universit\'e Libre de Bruxelles \\
  Service de Chimie-Physique and \\
 Center for Nonlinear Phenomena and
Complex Systems   C.P. 231 \\ 1050
Bruxelles, Belgium}
\date{\today}
\maketitle

\begin{abstract} 
We investigate the propagation of zero  sound in two-species
interpenetrating Bose-Einstein condensates. In very elongated
clouds, this propagation is shown to be essentially one-dimensional.
We also present 1D numerical experiments that exhibit the
existence of two different sound pulses with velocities
consistent with the theoretical predictions.
  \end{abstract} 

\indent
The dynamical behaviour of one-component Bose-Einstein condensates (BEC)
has been the subject of numerous experimental and theoretical studies
\cite{b1,b2}. In particular, measurements of the velocity of sound pulses
propagating in a cloud of \mbox{$^{23}Na$} atoms have been shown to agree with the prediction
of Bogoliubov theory \cite{b3}. On the other hand, Myatt {\it et al}. have
reported the observation of a double condensate created by thermal contact
between two different hyperfine states of \mbox{$^{87}Rb$} \cite{b4}.
This technique of
sympathetic cooling combined with the design of an optical dipole trap
\cite{b5} opens up the way to the realization of different multi-component
condensates. Several theoretical works have already been devoted to the
study of the ground state properties of two-species Bose-Einstein condensates
(TBEC) \cite{b6} as well as to the determination of the frequency of the
lowest collective excitations which correspond to global oscillations of
the clouds \cite{b7,b8,b9}.

\indent
In this paper, we study the propagation of zero sound in large two species
condensates. Such systems exhibit at long wavelengths two phonon branches.
As in the case of pure condensates \cite{b10,b11}, we show that in
long cigar-shaped clouds, this propagation takes place essentially
along the long axis. The sound velocities then take a simple
form that lends itself easily to numerical and experimental verifications.

\indent
Ultracold and dilute TBEC are well described by coupled Gross-Pitaevskii (GP)
equations for the macroscopic wavefunctions:

\begin{eqnarray}
    i\hbar \partial_t \Psi_1 = [
			-\frac{\hbar^2}{2m_1}{\underline \nabla}^2
			+ V_{e1} + U_{11}|\Psi_1|^2
			+ U_{12}|\Psi_2|^2 ] \Psi_1, 
			          \label{eq1}
\end{eqnarray}

\begin{eqnarray}
    i\hbar \partial_t \Psi_2 = [
			-\frac{\hbar^2}{2m_2}{\underline \nabla}^2
			+ V_{e2} + U_{22}|\Psi_2|^2
			+ U_{12}|\Psi_1|^2 ] \Psi_2, 
			          \label{eq2}
\end{eqnarray}

\noindent
together with the normalization conditions

\begin{eqnarray}
 \int d{ \underline r} |\Psi_i|^2 = N_i,                 \label{eq3}
\end{eqnarray}

\noindent
where the numbers of particles are conserved.

\indent
In these systems, the low energy binary collisions can be characterized by
the s-wave scattering lengths $a_{ij}.$ The interactions strengths are
given by

\begin{eqnarray}
 U_{ij} = 2 \pi \hbar^2 a_{ij} [ \frac{1}{m_i} + \frac{1}{m_j} ],
					\label{eq4}
\end{eqnarray}

\noindent
$U_{ij}$ is positive when the interaction is repulsive and negative for attractive
potentials. Few works have been devoted to the determination of the scattering
lengths between unlike alcalis. In the following, we restrict ourselves to the
case $U_{11}>U_{22}>0.$ The second terms in Eqs. (\ref{eq1}) and (\ref{eq2})
are the anisotropic
harmonic confining potentials:

\begin{eqnarray}
	V_{ei}= \frac{m_i \omega_{\perp i}^2} {2} [x^2+y^2+ \lambda^2 z^2],
					\label{eq5}
\end{eqnarray}

\noindent
where $\frac{m_1 \omega_{\perp 1}^2} {m_2 \omega_{\perp 2}^2}=\frac{g_1}{g_2}$
\cite{b12}; $g_i$ is the Land\'e $g$
factor. We only consider mixtures of alkali atoms for which $g_1=g_2$ and
Eq. (5) then becomes

\begin{eqnarray}
	V_e ({\underline r})= \frac{\kappa}{2} R^2; \ \ \ \ \
	R^2=x^2+y^2+ \lambda^2 z^2,
					\label{eq6}
\end{eqnarray}

\noindent
where $\kappa=m_i \omega^2_{\perp i}$.
   
\indent
As in the MIT experiments \cite{b3}, we investigate the case of very anisotropic
cigar-shaped potentials for which the asymmetry parameter
$\lambda = \frac{\omega_{\perp}}{\omega_{||}}$
is
very small. The long axis lies in the $z$ direction.

\indent
In order to discuss the behaviour of collective excitations, it is convenient
 to introduce the Madelung transformation:

\begin{eqnarray}
 \Psi_i = \sqrt{n_i} e^{i \phi_i},
 \label{eq7}
\end{eqnarray}

\noindent
where $n_i$ is the number density of the species $i$ and, in absence of vortices,
the phases $\phi_i$ play the role of potential for the two irrotational
velocities ${\underline v}_i = \frac{\hbar}{m_i} {\underline \nabla} \phi_i.$
In terms of these variables, the coupled GP equations
take the following simple form:

\begin{eqnarray}
	\frac{\partial n_i}{\partial t} + {\underline \nabla} . (n_i {\underline v}_i) = 0,
	\label{eq8}
\end{eqnarray}

\begin{eqnarray}
	m_i \frac{\partial {\underline v}_i}{\partial t} +
	{\underline \nabla} \{ V_e({\underline r}) +
	\sum_j U_{ij} n_j + \frac{m_i v_i^2}{2} + p_{qi} \} = 0,
	\label{eq9}
\end{eqnarray}

\noindent
where

\begin{eqnarray}
	p_{qi} = - \frac{\hbar^2}{2 m_i \sqrt{n_i}} {\underline \nabla}^2 (\sqrt{n_i})
	\label{eq10}
\end{eqnarray}

\noindent
is the quantum pressure.

\indent
When the numbers of atoms are sufficiently large, the density profiles become
smooth and the kinetic energy is small with respect to the interaction and
potential energies. It only takes a significant contribution at the boundaries.
When this term is neglected ($p_{qi}=0$), one can obtain simple analytical expressions
 for the density distributions in the atomic clouds (Thomas-Fermi approximation);
 in the case where $U_{11}>U_{22}$ we get

\begin{eqnarray}
	n_{T2} ({\underline r}) = \frac{\kappa \delta U_{11}}{2 \Delta}
		[R_2^2 - R^2]; \ \ \ {\textstyle for} \  R < R_2,
		\label{eq11}
\end{eqnarray}

\noindent
and similarly for $n_1$:

\begin{eqnarray}
	n_{T1} ({\underline r}) = \frac{\kappa}{2 U_{11}} [R_1^2 - R_2^2]
		+\frac{\kappa \delta U_{22}}{2 \Delta}
		[R_2^2 - R^2]; \ \ \ {\textstyle for} \  R < R_2,
		\label{eq12}
\end{eqnarray}

\noindent
and

\begin{eqnarray}
	n_{T1} ({\underline r}) = \frac{\kappa}{2 U_{11}} [R_1^2 - R^2];
		\ \ \ {\textstyle for} \ R_2 < R <R_1,
		\label{eq13}
\end{eqnarray}

\noindent
and the profiles vanish elsewhere.

\indent
In these equations, \mbox{$\delta U_{ii}=U_{ii} - U_{12}$} and
\mbox{$\Delta = U_{11} U_{22} - U_{12}^2$}. The radii \mbox{$R_1$}
and \mbox{$R_2$}
of
the clouds in the transverse plane at \mbox{$z=0$} can be determined
with the use of
Eq. (\ref{eq3}).

\indent
In free space \mbox{($V_e =0$),} the binary mixture is thermodynamically
stable with respect
to phase separation if

\begin{eqnarray}
	U_{11} U_{22} - U_{12}^2 > 0; \ \ \ \ \ \ \    \Delta >0.
	\label{eq14}
\end{eqnarray}

\indent
In the presence of the confining potential the two coupled clouds
interpenetrate
like miscible fluids as long as \mbox{$\delta U_{22} > 0$}
and the species with the largest self-interaction term (eg. 1) is
more spread
(cf. the bottom image in Fig. (1)). From Eq. (\ref{eq12}) one can see
that the species 1 experience
an inverted harmonic potential near the origin \mbox{($R<R_2$)} when
\mbox{$\delta U_{22} < 0$.} The peak density of 1 does not coincide
anymore with
that of the other species, but it takes place at \mbox{$R=R_2$.}
If the self-interaction \mbox{$\delta U_{22}$} is further decreased
the species 1
is expelled from the center of the trap and form a shell at the periphery
\cite{b6,b14}.
In the following, we consider the case of two interpenetrating
condensates, i.e.
$\Delta > 0$ and $\delta U_{22} > 0$.

\indent
It is useful to first recall the main formulas describing the
propagation of
sound in absence of external potential. The densities $n_{0i}$ of
the uniform
condensate are then equal to the peak densities at ${\underline r} = 0$
obtained
from Eqs. (\ref{eq11}) and (\ref{eq12}).  The frequencies $\omega_\kappa$
of the perturbations about
this uniform condensate and proportional to
$e^{i[\omega_\kappa t - {\underline \kappa . \underline r}]}$
are given by the following dispersion relations:

\begin{eqnarray}
	\omega_{\kappa \pm} = \frac{\kappa}{2}
		\{ [{\tilde U}_{11} + {\tilde U}_{22}] \pm
		[( {\tilde U}_{11} - {\tilde U}_{22})^2
		+ 4 {\tilde U}_{12}^2 ] ^{1/2} \} ^{1/2},
		\label{eq15}
\end{eqnarray}

\noindent
where

\begin{eqnarray}
	{\tilde U}_{ii} = \frac{\hbar^2 \kappa^2}{2 m_i^2}
		+ \frac{2 U_{ii} n_{0i}}{m_i},
		\nonumber
\end{eqnarray}

\noindent
and

\begin{eqnarray}
	{\tilde U}_{12} = 2 U_{12}
		 \{ \frac{n_{01} n_{02} }{m_1 m_2} \} ^{1/2}.
		 \nonumber
\end{eqnarray}

\indent
Taking $U_{12}=0$ and $U_{11}=0$ or $U_{22}=0$ in Eq. (\ref{eq15}),
one recovers the standard
Bogoliubov dispersion
relations for the pure condensates of each species. In a TBEC, the
mode corresponding to composition inhomogeneities is propagative and not
diffusive as in the case of normal binary mixtures. Both frequencies in
 Eq. (\ref{eq15}) are gapless and in the long wavelength limit
($\kappa \rightarrow 0$), they yield two phonon-like branches:

\begin{eqnarray}
	\omega_{\pm} (\kappa \rightarrow 0) \rightarrow c_{\pm} \kappa,
	\label{eq16}
\end{eqnarray}

\noindent
where the corresponding sound velocities are given by \cite{b15}

\begin{eqnarray}
	c_{\pm} = \frac{1}{\sqrt{2}} \{ ({c_{01}}^2 +  {c_{02}}^2)
		\pm  [({c_{01}}^2 -  {c_{02}}^2)^2
		+ \frac{4 U_{12}^2 n_{01} n_{02}}{m_1 m_2} ]^{1/2}
		\}^{1/2},
		\label{eq17}
\end{eqnarray}

\noindent
where $c_{0i} = (\frac{U_{ii} n_{0i}}{m_i})^{1/2}$ is the Bogoliubov
expression for the sound velocity
of a pure condensate of the species $i$. These collisionless modes correspond
to the Goldstone modes which proceed from the spontaneous breakdown of the
gauge symmetries ${\bar \Psi_i} \rightarrow e^{i \phi_i} \Psi_i$
in the composite condensate. In the mode associated
to the frequency $\omega_+$, the two components interfere
constructively whereas
this interference is destructive in the case of $\omega_-$. As a result,
the corresponding sound velocity $c_-$ can become imaginary signalling a
long wavelength instability leading to phase separation when
$\Delta < 0$ \cite{b16}.
This condition is consistent with the thermodynamic criterion
(cf. Eq. (\ref{eq14})).

\indent
In the presence of the confining potential, the linearized equation for the
perturbations $\delta n_i({\underline r},t)$ about the stationary
profiles $n_{ei}$ take
the following form after elimination of the velocity fields:

\begin{eqnarray}
      \frac{\partial^2 \delta n_1}{\partial t^2} = {\underline \nabla}
	. \{ \frac{U_{11} n_{e1}}{m_1} {\underline \nabla} \delta n_1 \}
	+ {\underline \nabla}
        . \{ \frac{U_{12} n_{e1}}{m_1} {\underline \nabla} \delta n_2 \},
	\label{eq18}
\end{eqnarray}

\begin{eqnarray}
      \frac{\partial^2 \delta n_2}{\partial t^2} = {\underline \nabla}
	. \{ \frac{U_{12} n_{e2}}{m_2} {\underline \nabla} \delta n_1 \}
	+ {\underline \nabla}
        . \{ \frac{U_{22} n_{e2}}{m_2} {\underline \nabla} \delta n_2 \}.
	\label{eq19}
\end{eqnarray}

\indent
When the densities are sufficiently high, the profiles $n_{ei}$ are well
described by the Thomas-Fermi approximation (Eqs. (\ref{eq11})-(\ref{eq13})).
This dynamics is greatly simplified in very elongated traps such as those
used in the MIT experiments on pure BECs \cite{b3}. Since the transverse
dimensions are smaller than the longitudinal ones, the density
profiles across the trap
quickly reach their equilibrium values. After this short transient, the
propagation is essentially one-dimensional and it can be described by a
local density perturbation that depends only on the longitudinal dimension
$\delta n_i (z;t)$ \cite{b10,b11}. Moreover, at the lowest order
in the smallness parameter
$\lambda$, one may neglect the variation of the equilibrium
profiles $n_{ei}$ in the $z$ direction (cylindrical geomety).
Eqs. ({\ref{eq18}) and ({\ref{eq19}) then become after integration over the
cross section of the cloud of species 1 perpendicular to the direction
of propagation:

\begin{eqnarray}
    \frac{\partial^2 \delta n_1}{\partial t^2} =
	\frac{U_{11} < n_{e1} >}{m_1}
	(\frac{\partial^2 \delta n_1}{\partial z^2})
	+ \frac{U_{12} < n_{e1} >}{m_1}
	(\frac{\partial^2 \delta n_2}{\partial z^2}),
	\label{eq20}
\end{eqnarray}
	
\begin{eqnarray}
    \frac{\partial^2 \delta n_2}{\partial t^2} =
	\frac{U_{12} < n_{e2} >}{m_2}
	(\frac{\partial^2 \delta n_1}{\partial z^2})
	+ \frac{U_{22} < n_{e2} >}{m_2}
	(\frac{\partial^2 \delta n_2}{\partial z^2}),
	\label{eq21}
\end{eqnarray}

\noindent
where

\begin{eqnarray}
	n_{ei}=\frac{2}{R_i^2} \int_{0}^{R_1} r dr n_{ei}({\underline r}).
\end{eqnarray}

\indent
As a result, sound propagation in such very anisotropic systems can be
described by coupled one-dimensional wave equations. The corresponding
sound velocities are still given by Eq. (\ref{eq17}), but the densities
of the uniform condensates are now replaced by the average densities
as over the large transverse section of the cloud.

\indent
	The effect of a density inhomogeneity on such two-species
Bose-Einstein condensates can thus be captured by one-dimensional numerical
experiments. The initial conditions were created by imaginary time
integration of the coupled GP equations in the presence of an harmonic
confining potential. Once the initial condition was
obtained, we perturbated the condensate thus causing the appearance of
inhomogeneities in the center of our mixtures. This perturbation was
created by instantaneously adding to the initial external potential a
Gaussian tip in the middle of the confining region.
It mimics the laser beam that is focused in the experiments in the
center of the trap to create a repulsive dipole force \cite{b3}.
A variant of this experiment
consists of creating the initial condensate with an external potential
that includes the Gaussian barrier and eliminating it instantaneously
at time $t=0$. Both constructions give qualitatively the same results.
	We show one  such integration on Fig. (1).
The parameters were chosen to guarantee that the mixture is thermodynamically
stable and the clouds interpenetrate.
For these parameters, the Thomas-Fermi approximation permits to
calculate the extension of the condensates which are  
$462$ and $570$ $\mu m$. The
numerically created condensates have widths of approximately $450$ and
$550$ $\mu m$, indicating a fairly good agreement between both aproaches.
	At $t=0$ we have switched on the above mentioned Gaussian barrier
to perturbate  the system.
This  generates two pulses which propagate symmetrically
outwards.
In the beginning one sees only one hump (on each side of the mixture). 
However, as time increases, one can distinguish two humps one of which
outdistances the other. The position of these density minima varies
linearly with time and the speeds of propagation can thus be extracted
from the simulations.
They are of approximately
$10.4$ and $8.1$ cm/s. When we calculate them by using the Thomas-Fermi
 peak densities
into the expressions for the sound speed,
we find $11.7$ and $9.4$ cm/s. Again there is
a fairly good agreement between the integration of the coupled GP equations 
 and Thomas-Fermi approximation results.

\acknowledgements  D. L. received support from the
CNPq/Brazil and P. B. and G. D. from the FNRS
(Belgium). D. L. would like to thanks the hospitality at the Center
for Nonlinear Phenomena and Complex Systems (ULB).\\

\noindent Figure 1.
Numerical integration of the coupled GP equations.
The bottom image corresponds to the initial condition.
The upper images show the development in time (increasing upwards
and given in $\mu s$)
after the perturbation has been applied.
On the y-axis we plot the
density of these condensates in $\mu m^{-3}$, while  the abscissa represents their spatial
extension in
$\mu m$. 
The
dashed line corresponds to condensate of particles 1. The dot-dashed line
corresponds to condensate of particles 2. 
The sum of the densities is plotted as a solid line. For
this figure the masses of both species are $10^{-26} Kg$. Each species
contributes with $5 \ 10^6$ particles, and the frequency of the harmonic
oscillator is equal to $300 Hz$. Particles of type 1 and 2 present s-wave
scattering length of the $1.5 \ 10^{-3}$ and $8.0 \ 10^{-4}$ $\mu m$,
while the interspecies
length is of $1.0 \ 10^{-4}$ $\mu m$. At $t=0$ a Gaussian barrier was
switched on to perturbate the mixture.


{\small \begin{thebibliography}{Dillo83}

\bibitem{b1} F. Dalfovo {\it et al.}, cond-mat/9806038 (1998) and references
           therein.

\bibitem{b2} A. S. Parkins and D. F. Walls, Phys. Reports
	{\bf 303}, 4 (1998).

\bibitem{b3} M. R. Andrews {\it et al.}, Phys. Rev. Lett. {\bf 79},
         553 (1997); Phys. Rev. Lett. {\bf 80}, 2967 (1998).

\bibitem{b4} C. J. Myatt {\it et al.}, Phys. Rev. Lett. {\bf 78},
	586 (1997).

\bibitem{b5} D. M. Stamper-Kurn {\it et al.}, Phys. Rev. Lett.
	{\bf 80}, 2027 (1998).

\bibitem{b6} R. Ejnisman {\it et al.},  Optics Express {\bf 2}, 330
          (1998) and references therein.

\bibitem{b7} Th. Busch {\it et al.}, Phys. Rev. A {\bf 56}, 2978 (1997).

\bibitem{b8} H. Pu and N. P. Bigelow,  Phys. Rev. Lett. {\bf 80},
	1134 (1998).

\bibitem{b9} B. D. Esry and C. H. Greene,  Phys. Rev. A {\bf 57},
	1265 (1998).

\bibitem{b10} E. Zaremba,  Phys. Rev. A {\bf 57}, 518 (1998).

\bibitem{b11} G. M. Kavoulakis and C. J. Pethick,  Phys. Rev. 
            A {\bf 58}, 1563 (1998).


\bibitem{b12} T. L. Ho and V. B. Shenoy,  Phys. Rev. Lett.
             {\bf 77}, 2595 (1996).

\bibitem{b13} D. M. Larsen,  Ann. of Phys. {\bf 24}, 89 (1963).

\bibitem{b14} H. Pu and N. Bigelow,  Phys. Rev. Lett. {\bf 80},
	1130 (1998).

\bibitem{b15} R. Graham and D. Walls,  Phys. Rev. A {\bf 57},
	484 (1998).

\bibitem{b16} E. V. Goldstein and P. Meystre,  Phys. Rev. A
          {\bf 55}, 2935 (1997).
	

\end{thebibliography} }
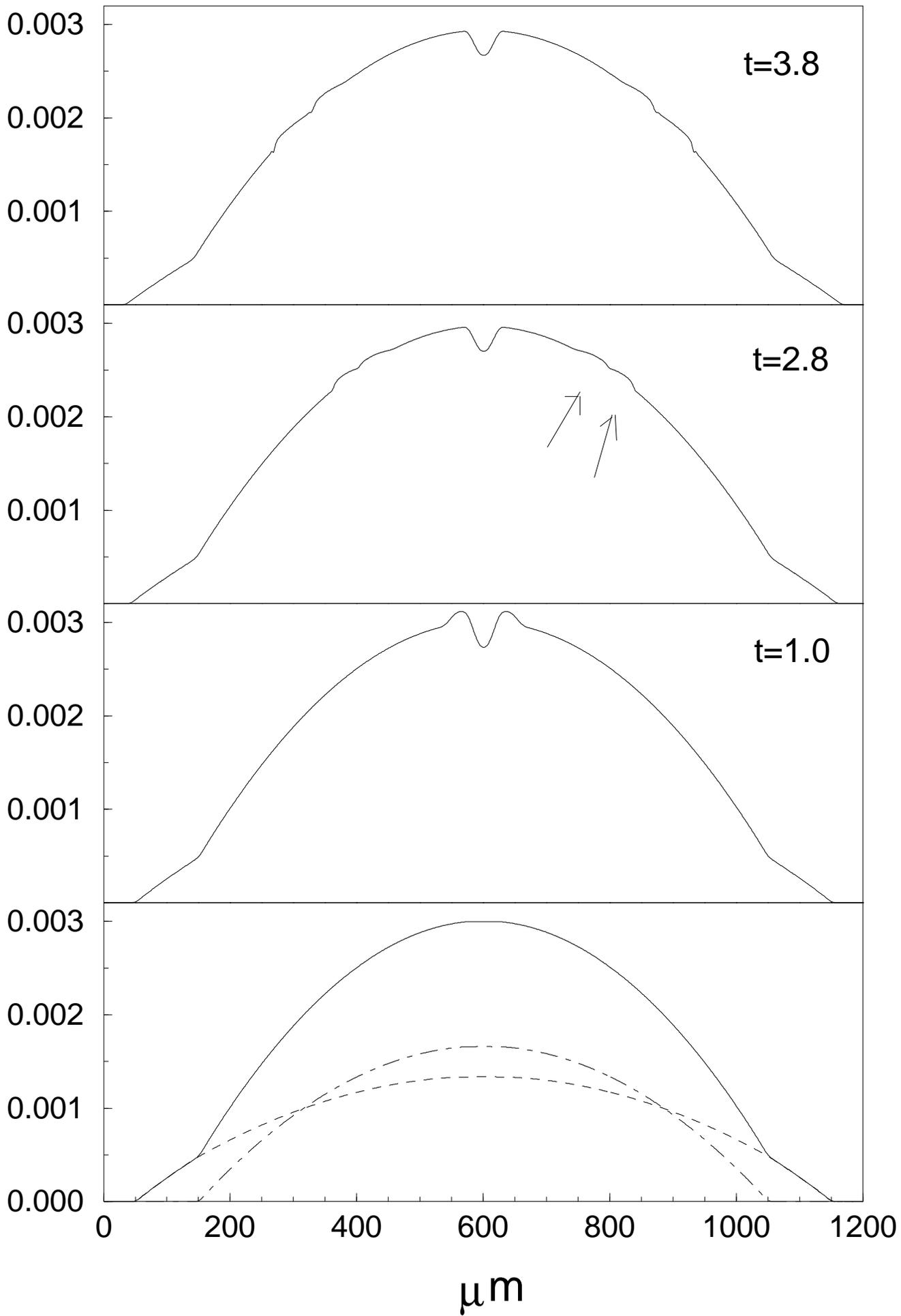
\end{document}